\documentstyle[aps,prb,psfig]{revtex}
\begin{document}

\title{Specific heat and thermal conductivity in Sr$_2$RuO$_4$   for
rotating in-plane magnetic field}
\author{L. Tewordt and D. Fay}
\address{I. Institut f\"ur Theoretische Physik,
Universit\"at Hamburg, Jungiusstr. 9, 20355 Hamburg, 
Germany}
\date{\today}
\maketitle
\begin{abstract}
      We calculate the density of states $N$ and the the thermal 
conductivity $\kappa\,$ in layered superconductors for rotating in-plane
magnetic field by using approximate analytical expressions for Abrikosov's
vortex lattice state. For a gap with d$_{xy}\,$- orbital symmetry we find that the 
differences $\Delta N$ and $\Delta\kappa$ between field directions parallel
to the antinodes and parallel to the nodes of the gap are positive for lower 
fields and change sign for higher fields near $H_{c2}\,$. For increasing 
impurity scattering, $\Delta N$ and $\Delta\kappa$ decrease due to a 
broadening of certain peaks in the angle-resolved density of
states. The frequency dependencies of the amplitudes 
$\Delta N$ and $\Delta\kappa$ exhibit similar crossover behavior at low
frequencies which give rise to a sign change of $\Delta\kappa$ at 
low $T/T_c\,$.  We conclude that variations of the specific heat and
thermal conductivity can be observed only in the clean and low
temperature limits.
\end{abstract}
\pacs{74.25.Fy, 74.72.-h, 74.25.Op, 74.70.Pq}
\vspace{0.5in}
       Recently fourfold oscillations of the specific heat have been observed
for rotating in-plane magnetic field in pure Sr$_2$RuO$_4$ at very low 
temperatures. \cite{Deguchi} These data have been analyzed on the basis
of the Bogoliubov-de Gennes (BdG) equations and the Pesch-approximation
for the vortex state with the result that the spin-triplet superconductor 
Sr$_2$RuO$_4$  has a gap with nodes in the directions of the a and b 
axes. \cite{Udagawa} In an earlier measurement of the thermal
conductivity $\kappa$ no appreciable angular variation was observed from 
which it was concluded that no vertical line nodes exist in 
Sr$_2$RuO$_4\,$. \cite{Izawa} However, the amplitude of the oscillation 
of $\kappa$ for a state with vertical gap nodes was found to be strongly
temperature dependent and to change sign for increasing temperature.
\cite{TF1} This explains the failure to observe an angular variation of 
$\kappa$ because the measuring temperature in the experiment of 
Ref.~\onlinecite{Izawa} was too high.

      In the present paper we repeat the calculations of the density of states
$N$ and the thermal conductivity $\kappa$ for rotating in-plane field for a 
superconductor with d$_{xy}$ symmetry of the orbital part of the gap using
the analytical expressions of Pesch \cite{Pesch,Vek} based on the 
quasiclassical Green's functions which are easier to handle than the 
original expressions based on the Gorkov equations and Abrikosov's 
vortex state. \cite{BPT,TF1} Our main purpose is to investigate the effect of 
impurity scattering. This depends sensitively on the phase shift for impurity
scattering as has been shown in a recent paper for d-wave pairing
in fields parallel to the c-axis. \cite{TF2} We shall only present results for
the unitary limit of impurity scattering with phase shift $\pi/2$ because the
relevance of the unitary limit has been strongly suggested by the results
for the universal heat transport in Sr$_2$RuO$_4\,$. \cite{Suzuki}

      For simplicity we neglect the small c-axis dispersion of the cylindrical
Fermi surface and thus the c-axis component of the Fermi velocity. The
azimuthal angle of the quasiparticle momentum in the ab-plane with respect
to the a-axis is denoted by $\phi$ and and we use $\alpha$ for the direction
of the magnetic field in the ab-plane. Thus the component of the Fermi
velocity in the direction perpendicular to the magnetic field is given by
$v\sin(\phi - \alpha)\,$. For the momentum part of the order parameter we
assume d$_{xy}$ symmetry which, together with the field dependence of
the spatial average of the modulus of the Abrikosov vortex state, yields the
expression
\begin{equation}
|\Delta |^2 = \Delta^2_0 \,  [1-h^2]^{1/2}\sin^2(2\phi)
\label{Del}
\end{equation}
where $h=H/H_{c2}$ and $H$ is the spatial average of the field. The
Pesch-approximation now yields for the spatial average of the density of
states $N(\omega)\,$, normalized to the normal state density of states
$N_0\,$:
\begin{equation}
N(\omega)/N_0 = \mbox{Re}\left\{1+8|\Delta |^2 \,
[\Lambda/v\,\sin(\phi - \alpha)]^2 \, [1+i\sqrt{\pi}\,z\, w(z)]\right\}^{-1/2}\,;
\label{N}
\end{equation}
where
\begin{equation}
z=2[\omega+i\,\Sigma_{\mbox{i}}][\Lambda/v\,|\sin(\phi - \alpha)|\,]\:;
\qquad \Lambda=(2eH)^{-1/2}\: .
\label{z}
\end{equation}
Here $\Sigma_{\mbox{i}}$ is the self energy for impurity scattering which is
calculated self-consistently in the t-matrix approximation from the 
expressions given in Ref.~\onlinecite{TF2}. The thermal conductivity 
$\kappa$ in the BPT-approximation is given in Ref.~\onlinecite{TF1} and, 
in the equivalent Pesch-approximation in Ref.~\onlinecite{Vek}. The 
integrand of the $\omega$-integral in Ref.~\onlinecite{Vek} is proportional to
the density of states [Eq.(\ref{N})]  times the lifetime of the quasiparticles. 
This lifetime is equal to the reciprocal of the sum of the scattering rates due
to impurity and Andreev scattering which have been discussed at length in
Ref.~\onlinecite{TF2}.    

      In Fig.1 we show our results for $N(\omega=0)/N_0$ and 
$\kappa(T=0)/\kappa_n$ ($\kappa_n$ is the normal state thermal
conductivity) versus the reduced field $h=H/H_{c2}$ for fields along the
gap node ($\alpha = 0\,$, solid curves) and along the antinode
($\alpha = \pi/4\,$, dashed curves). The heat current for $\kappa$ is
along the antinode of the gap, $\phi = \pi/4\,$. The impurity scattering
self energy is calculated self-consistently for the unitary phase shift
limit and the reduced scattering rate $\delta = \Gamma/\Delta_0 = 0.01$ 
where $\Gamma=1/2\tau_n$ is the normal state scattering rate. This is the
right order of magnitude for the purest samples of Sr$_2$RuO$_4$ with
the highest $T_c\,$. \cite{Suzuki} One sees that the dashed curve for 
$N/N_0$ (antinode) lies above the solid curve (node) for $h$ from zero to
about the crossing field $h_c=0.63\,$ while, above $h_c\,$, it lies below
the solid curve. The difference $\Delta N$ between the two curves 
corresponds to the amplitude of the measured fourfold anisotropy of the
specific heat. \cite{Deguchi} This amplitude is about 6 percent of
$N(\alpha=0)$ at $h=0.3\,$. Our results agree approximately with the 
results obtained in Ref.~\onlinecite{Udagawa} by solving the BdG
equations and by using the Pesch-approximation. For $\kappa/\kappa_n$
(lower curves in Fig.1) the dashed curve (antinodal direction of the
field) lies above the solid curve (nodal direction of the field) for $h$ 
between 0 and about $h_c = 0.95$ where it crosses the solid curve.
As with the density of states, the difference $\Delta\kappa$ between the 
dashed and solid curve corresponds the amplitude of the fourfold 
anisotropy and amounts to about 25\% of $\kappa(\alpha=0)\,$. It has been
shown in Ref.~\onlinecite{TF1} that the amplitude $\Delta\kappa$ of the 
fourfold oscillation in rotating field decreases rapidly with increasing
temperature and reverses its sign at a very low temperature indicating
that the variations with field direction can only be observed at very low
temperatures.

      In addition to the temperature effect, impurity scattering also 
diminishes the amplitudes of the fourfold oscillations of the density of 
states and the thermal conductivity. An example is shown in Fig.2 where
we present our results for the unitary phase shift limit and a much larger 
scattering rate $\delta=0.1$ which is of the order of magnitude
corresponding to lower $T_C\,$'s in Sr$_2$RuO$_4\,$. \cite{Suzuki}
Comparison of Fig.2 with 
Fig.1 shows that, for the larger scattering rate, the crossing field 
$h_c\simeq 0.4$ for the curves of $N$ is much smaller and the relative 
amplitude $\Delta N/N(\alpha=0) \sim \,$2\% is also much smaller. The 
crossing field for $\kappa$ and the relative amplitude
$\Delta\kappa/\kappa(\alpha=0)$ are of the same orders of magnitude as
those for $N\,$.  

      In Ref.~\onlinecite{Udagawa} the reasons for the crossover behavior
of the density of states have been explained in detail by calculating the
angle-resolved density of states for low and high fields using the BdG 
equations and the Pesch-approximation. Similar angle-resolved 
densities of states $N(\phi)/N_0$ are shown in Fig.3 for a low field 
$h=0.2$ and nodal field direction ($\alpha=0\,$, solid curves) and 
antinodal field direction ($\alpha=\pi/4\,$, dashed curves). Note that, in the
latter case, the angle $\phi$ is measured from the field direction 
$\alpha=\pi/4\,$. One sees that, for $\alpha=0\,$, a single broad peak
occurs at $\phi=\pi/2$ while, for $\alpha=\pi/4\,$, two peaks occur at
$\phi=\pi/4$ and $\phi=3\pi/4\,$. This explains why in this field region the
total density of states $N(\alpha=\pi/4)$ is larger than $N(\alpha=0)$ 
because, for $\alpha=\pi/4\,$, all four nodes contribute to $N$ while, for
$\alpha=0\,$, two of the four nodes are parallel to the field and are thus
unable to contribute to $N\,$. In the high magnetic field region, on the
other hand, $N(\alpha)=0) > N(\alpha=\pi/4)$ because then the behavior
is determined by $N(\phi)$ for $\phi=0$ and $\phi=\pi\,$. 
Comparison of Fig.3(a), calculated for the unitary limit and $\delta=0.01\,$,
with Fig.3(b), calculated for $\delta=0.1\,$, shows that the main effect of
stronger impurity scattering is to broaden the peaks of the solid curve at
$\phi=0$ and $\phi=\pi$ and to enhance the minima at 
$\phi=\pi/4$ and $3\pi/4\,$. At the same time the minima of the dashed
curve $\phi=0$ and $\phi=\pi$ are enhanced. This effect of impurity
scattering is similar to the effect of finite c-axis dispersion of the 
cylindrical Fermi surface which was taken into account in 
Ref.~\onlinecite{Udagawa}. The total effect of the broadening of the peaks
and the enhancement of the minima in the angle-resolved density of 
states for low fields is the decrease of the amplitudes of variation of 
$N$ and $\kappa$ in rotating in-plane field as shown in Figs.1 and 2.

      Finally we consider the frequency dependence of $N(\omega)/N_0$
for nodal and antinodal field directions and its effect on the frequency
and temperature dependence of the thermal conductivity. In Fig.4(a) we
have plotted our results for $N(\omega)/N_0$ versus 
$\Omega=\omega/\Delta_0$ for a low field ($h=0.2$) for nodal
($\alpha=0\,$, solid curve) and antinodal ($\alpha=\pi/4\,$, dashed curve)
field directions. In Fig.4(b) we show the corresponding curves for
$\kappa(\omega)/\kappa_n$ which is the factor multiplying 
$(\omega/T)^2 \,\mbox{sech}^{2}(\omega/2T)$ in the normalized integral 
over $d(\omega/T)$ for $\kappa/\kappa_n\,$. \cite{Vek} This function of
$\Omega=\omega/\Delta_0$ yields approximately the temperature
dependence of $\kappa/\kappa_n$ through the relation 
$\Omega \simeq 2.4(T/\Delta_0)$ because the integrand is strongly 
peaked at $\omega/T=2.4\,$. The expression for $\kappa(\omega)$ is
given by $N(\omega)\tau(\omega)$ where the quasiparticle scattering rate
$1/2\tau(\omega)$ is equal to the sum of the impurity and Andreev 
scattering rates. \cite{TF2} The most interesting results in Figs.4(a) and 4(b)
are the crossovers of the dashed and solid curves occuring at very low
frequencies which correspond to a sign change of the amplitudes of the 
variations of fourfold symmetry with rotation angle. For $\kappa(\omega)$
this first crossover occurs at about $\Omega_c \simeq 0.1$ which means
that, at a temperature of about $T/T_c \simeq 0.1ß,$, the amplitude of the
variation with rotation angle vanishes or changes sign.It should be 
pointed out that for high fields the curves for $N(\omega)$ and 
$\kappa(\omega)$ look quite different. The solid curve for $N(\omega)$ 
starts out at a much higher value than the dashed curve and the coherence
peak for the solid curve vanishes. This is similar to the results of
Ref.~\onlinecite{Udagawa}.

      In sumary, we have calculated the density of states $N$ and the
thermal conductivity $\kappa$ for a gap with d$_{xy}\,$- orbital 
symmetry and magnetic fields in the directions 
$\mbox{\boldmath$H$}$ $||\,$  gap node ($\alpha=0\,$, solid curves) and
$\mbox{\boldmath$H$}$ $||\,$  gap antinode 
($\alpha=\pi/4\,$, dashed curves). For both calculations we have used
the approximate expressions of Pesch \cite{Pesch,Vek} for Abrikosov's
vortex lattice state which are easier to handle than the original
expressions of the BPT-approximation. \cite{BPT,TF1} Comparison of
these results with the results of the quasiclassical Eilenberger 
equations \cite{Dahm} and the BdG-equations \cite{Udagawa} has
shown that these analytical expressions provide very good
approximations over the whole field range from $H_{c2}$ down to
$H_{c1}\,$. We have included impurity scattering and used a phase
shift of $\pi/2$ because the unitary limit in Sr$_2$RuO$_4$ is strongly
suggested by universal heat transport. \cite{Suzuki}

    Our main result is that the density of states $N(\omega=0)$ and the
thermal conductivity $\kappa(T=0)/\kappa_n$ are somewhat larger for the
antinodal field direction ($\alpha=\pi/4$) than for the nodal field
direction ($\alpha=0$) in a field region $0\le h \le h_c$ ($h=H/H_{c2}$)
where $h_c$ denotes the field where the dashed curves cross the solid
curves (see Fig.1). Our results for $N$ agree essentially with the 
results obtained from the BdG equations and the Pesch-approximation. 
\cite{Udagawa} These results have led to the conclusion that the
observed minima and maxima of the specific heat for rotating in-plane
field in Sr$_2$RuO$_4$ are due to vertical line nodes in the directions
of the a- and b-axes. \cite{Deguchi} It turns out that the amplitudes 
$\Delta N = N(\alpha=\pi/4) - N(\alpha=0)$ and 
$\Delta \kappa = \kappa(\alpha=\pi/4) - \kappa(\alpha=0)$ and the 
crossover field $h_c$ decrease for increasing impurity scattering 
(see Fig.2). The reason can be seen from a comparison of the
angle-resolved density of states $N(\phi)/N_0$ in Figs.3(a) and 3(b). 
The most important contributions to the density of states 
[(see Eqs.(2) and (3)] are the terms $\Lambda/v\,|\sin(\phi - \alpha)|$ 
where $\Lambda=(2eH)^{-1/2}$ is the magnetic length, 
$v\,|\sin(\phi - \alpha)|$ is the component of the Fermi velocity 
perpendicular to $\mbox{\boldmath$H$}\,$, and $\phi$ and $\alpha$
are the azimuthal angles of the quasiparticle momentum and field 
direction, respectively. Asymptotic expansion of the $w$-function 
\cite{BPT} in Eq.(2) shows that in the limit $\phi \rightarrow \alpha$ the
density of states takes on the BCS form for $|\Delta(\phi)|^2$ [see
Eq.(1)]. For quasiparticles moving in the directions perpendicular to
the field ($\phi-\alpha = \pm\pi/2$) the analytic expression involving thr
$w$-function yields the full effect of the superfluid flow and the Andreev
scattering due to the complex order parameter of Abrikosov's vortex
lattice function. These directional terms, together with the angular
dependence $\Delta^2_0\sin^2(2\phi)$ of the gap function, yield quite
different angle-resolved densities of states $N(\phi)/N_0$ for
$\alpha=0$ and $\alpha=\pi/4$ shown in Fig.3. Comparison of
Figs.3(a) and 3(b) shows that the main effect of stronger impurity
scattering is to broaden the peaks of the solid curve at $\phi=0$
and $\phi = \pi\,$. It is interesting that this effect is quite similar to the
effect of a finite c-axis dispersion of the cylindrical Fermi surface 
which has been taken into account in the analytic Pesch-expression
in Ref.~\onlinecite{Udagawa}. Finally, we have investigated the finite
temperature effect on the amplitude $\Delta\kappa$ for the variation with
rotating in-plane field by calculating $N(\omega)/N_0$ and
$\kappa(\omega)/\kappa_n$ for the field directions $\alpha=0$ and
$\alpha=\pi/4$ (see Figs.4(a) and 4(b) for low field $h=0.2$ and small
impurity scattering $\delta=0.01\,$). The large differences between 
$N(\omega)$ and $\kappa(\omega)$ arise from the quasiparticle 
lifetime $\tau(\omega)$ in the latter expression which is given by the
reciprocal of the sum of scattering rates due to impurity and Andreev
scattering. The impurity scattering rate is calculated self-consistently
in the unitary limit of the t-matrix approximation, and the Andreev
scattering rate arises from the imaginary part of the self energy 
proportional to $\Delta(r_1)\Delta^{\ast}(r_2)G(r_1-r_2,-\omega)\,$,
where $\Delta(r)$ is Abrikosov's vortex lattice order parameter 
\cite{TF2} and $G$ is the hole propagator. The crossover of the 
dashed and solid curves at a low
frequency $\omega \simeq 0.1\Delta_0$ in Fig.4(b) leads to a 
corresponding crossover in the temperature dependencies of $\kappa$
at about $T/T_c \simeq 0.1$ showing that the amplitude of the variation
with field rotation vanishes and changes sign at this temperature. 
Similar results have already been obtained in Ref.~\onlinecite{TF1}
using the BPT-approximation.

      In conclusion we can say that the amplitudes of the variations of the
specific heat and thermal conductivity in rotating in-plane field for a gap
with vertical line nodes decrease rapidly with increasing impurity
scattering and increasing temperature.
\newpage
\newpage
\begin{figure}
\centerline{\psfig{file=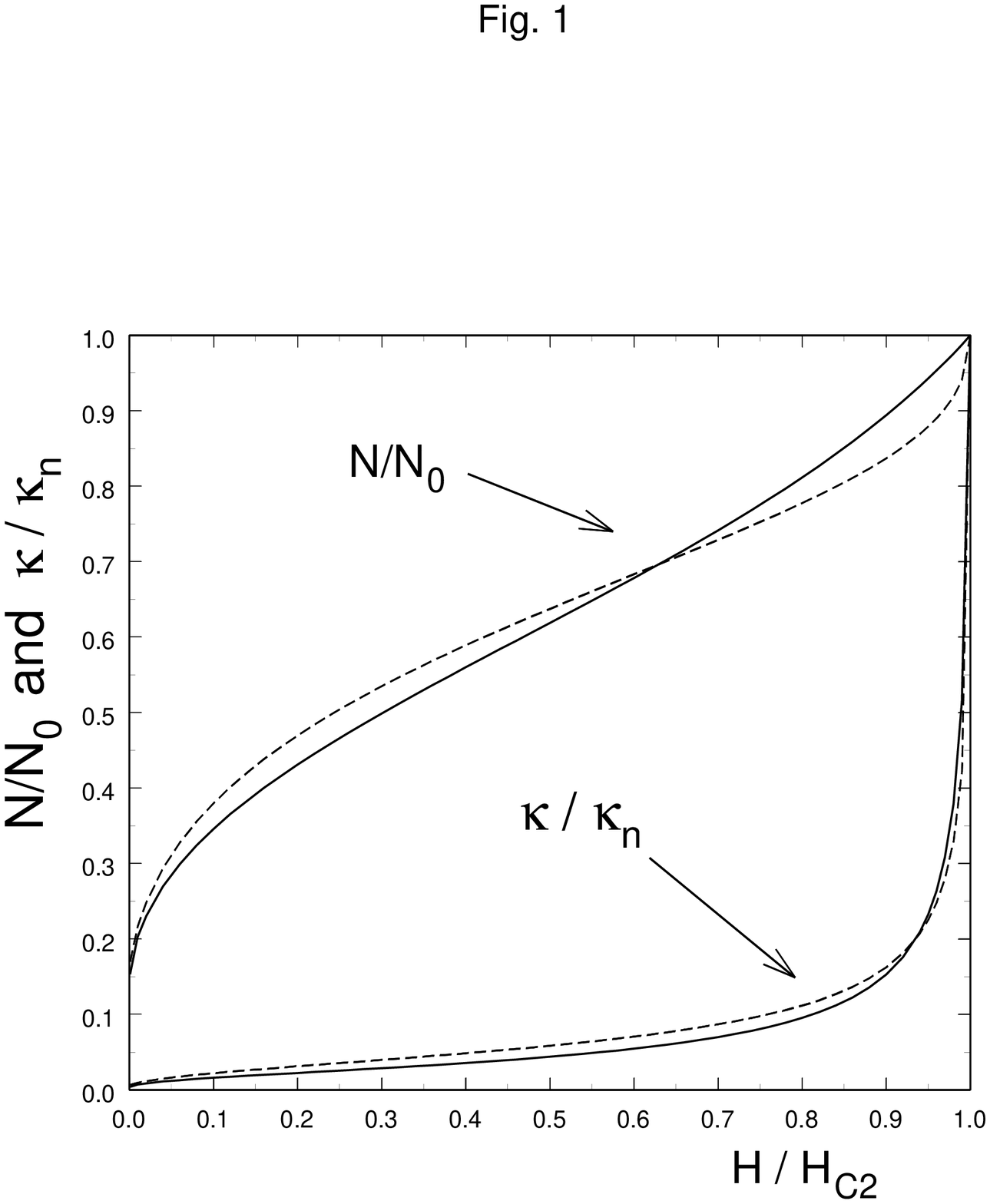,width=18cm,angle=0}}
\vskip -1cm
\caption{1) Density of states, $N(\omega=0)/N_0\,$, and thermal conductivity,
$\kappa/\kappa_n (T\rightarrow0)\,$, versus $h=H/H_{c2}$ for in-plane
field direction parallel to the gap node ($\alpha=0\,$, solid curves) and
parallel to the antinode ($\alpha=\pi/4\,$, dashed curves). The reduced
impurity scattering rate is $\delta=\Gamma/\Delta_0 = 0.01$ with phase
shift in the unitary limit.}
\label{fig1}
\end{figure}
\begin{figure}
\centerline{\psfig{file=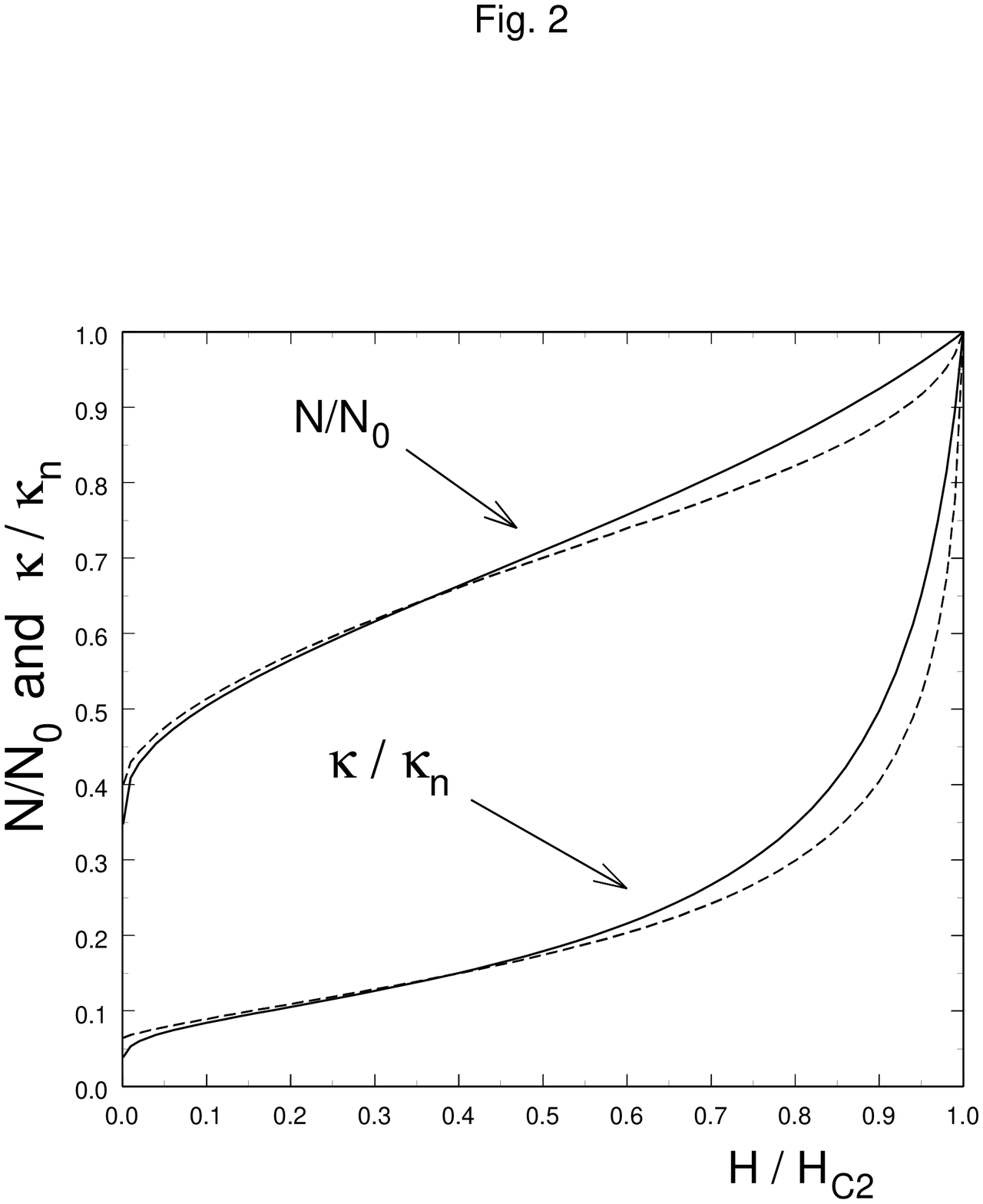,width=18cm,angle=0}}
\vskip -1cm
\caption{2) The same as Fig.1 but with $\delta=0.1\,$.}
\label{fig2}
\end{figure}
\begin{figure}
\centerline{\psfig{file=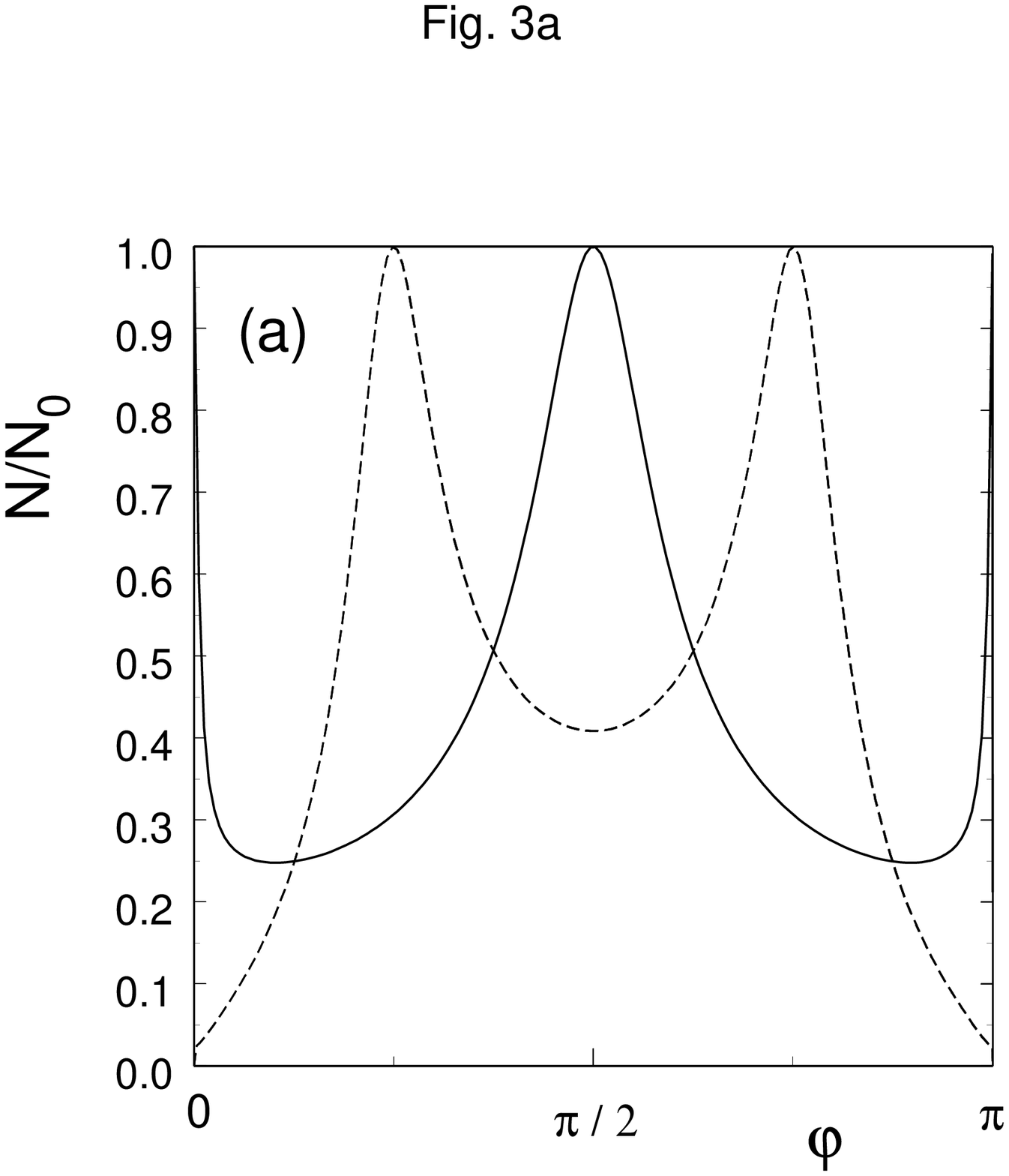,width=18cm,angle=0}}
\vskip -1cm
\caption{3a) Angle-resolved density of states, $N(\phi)/N_0\,$ for $h=0.2\,$. Solid
curves are for nodal field direction ($\alpha=0$) and dashed curves for
antinodal direction ($\alpha=\pi/4$). The angle $\phi$ is measured from
the field direction. Impurity scattering with $\delta=0.01$ in the unitary
limit.}
\label{fig3a}
\end{figure}
\begin{figure}
\centerline{\psfig{file=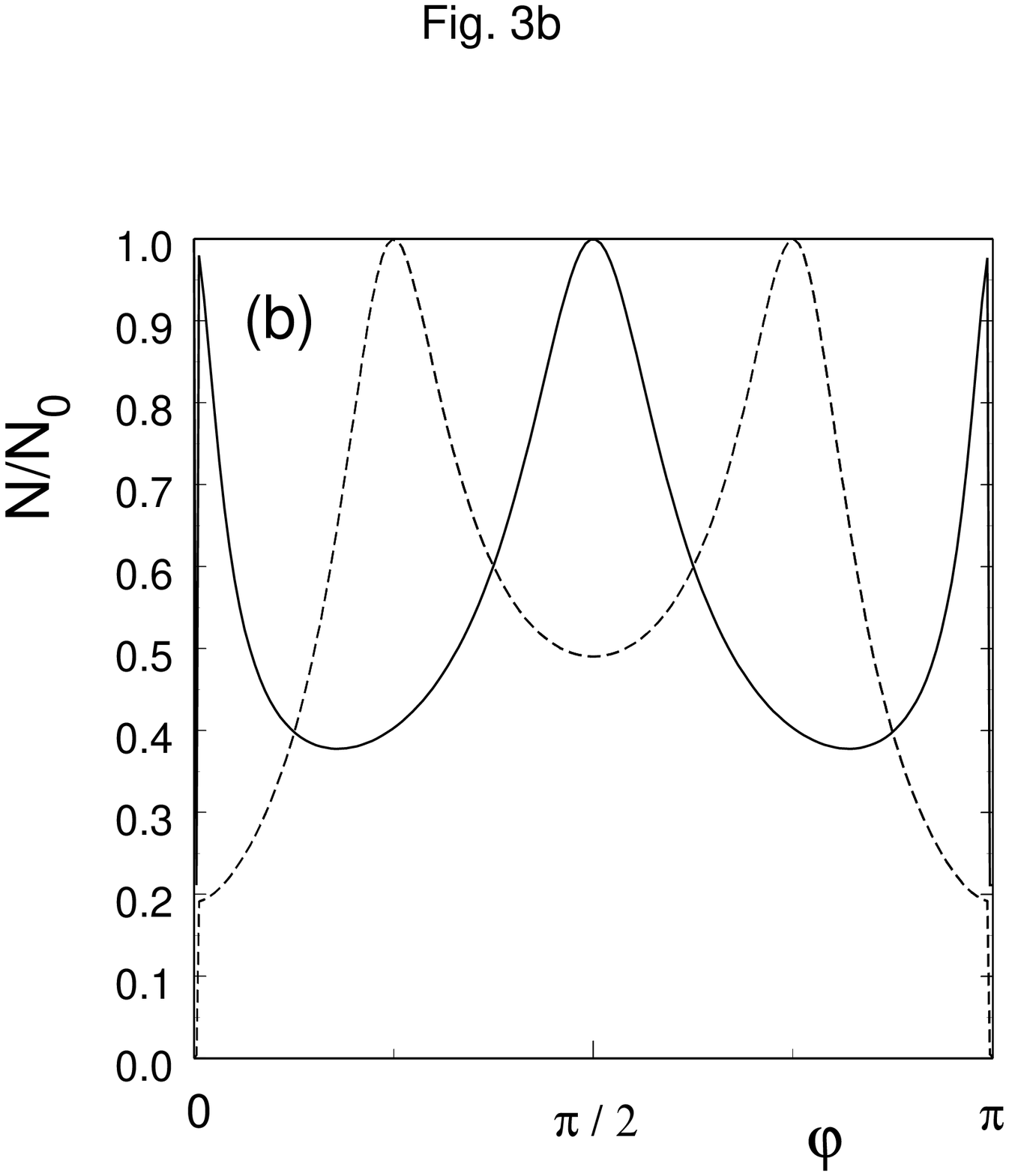,width=18cm,angle=0}}
\vskip -1cm
\caption{3b) Angle-resolved density of states, $N(\phi)/N_0\,$ for $h=0.2\,$. Solid
curves are for nodal field direction ($\alpha=0$) and dashed curves for
antinodal direction ($\alpha=\pi/4$). The angle $\phi$ is measured from
the field direction. Impurity scattering with $\delta=0.1$ in the unitary
limit.}
\label{fig3b}
\end{figure}
\begin{figure}
\centerline{\psfig{file=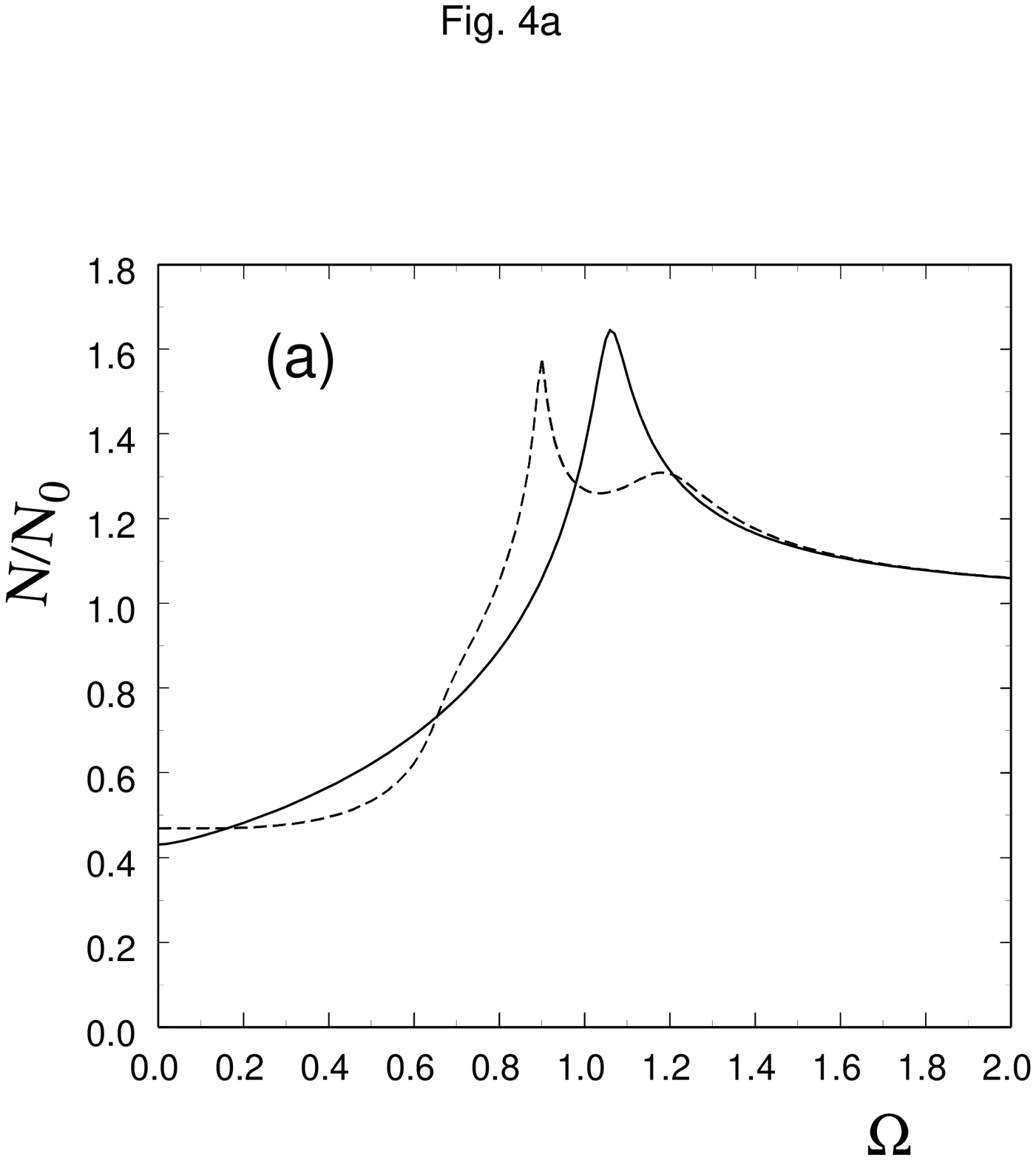,width=18cm,angle=0}}
\vskip -1cm
\caption{4a) Density of states $N(\omega)/N_0$ versus $\Omega=\omega/\Delta_0$
for $h=0.2\,$, $\delta=0.01\,$, and field directions $\alpha=0$ (solid curve)
and $\alpha=\pi/4$ (dashed curve).}
\label{fig4a}
\end{figure}
\begin{figure}
\centerline{\psfig{file=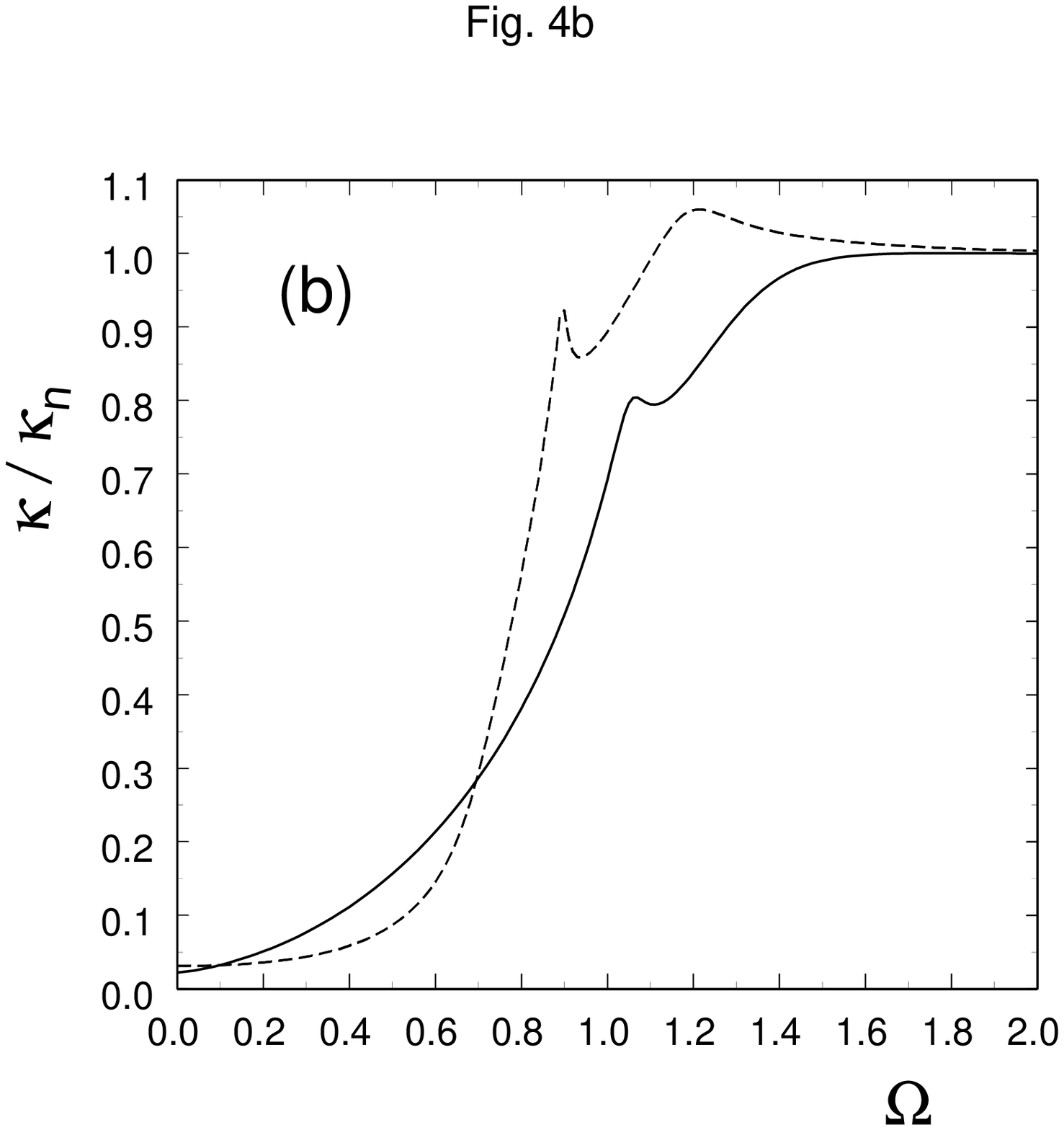,width=18cm,angle=0}}
\vskip -1cm
\caption{4b) Thermal conductivity 
$\kappa(\omega)/\kappa_n$ versus $\Omega=\omega/\Delta_0$
for $h=0.2\,$, $\delta=0.01\,$, and field directions $\alpha=0$ (solid curve)
and $\alpha=\pi/4$ (dashed curve). The temperature gradient is in the 
direction of the antinode.}
\label{fig4b}
\end{figure}
\end{document}